\documentclass[a4paper,pra,aps,12pt,nofootinbib]{revtex4}

\usepackage[T1]{fontenc}
\usepackage[utf8]{inputenc}
\usepackage{units}
\usepackage{textcomp}
\usepackage{amsmath}    
\usepackage{graphicx}   
\usepackage{verbatim}   
\usepackage{color}      
\usepackage{subfigure}  
\usepackage{hyperref}   
\hypersetup{citebordercolor=.2 .8 .2,urlbordercolor=0 .75 1}
\usepackage{nicefrac}
\usepackage{amsfonts}
\usepackage{amsthm}
\usepackage{amssymb}
\usepackage{booktabs}
\usepackage{braket}
\usepackage{fancyhdr}
\usepackage{moreverb}
\usepackage[brazil]{babel}

\makeatletter
\renewcommand*{\email}[1][E-mail: ]{\begingroup\sanitize@url\@email{#1}}
\makeatother

\begin{document}

	\title{Classical gauge theories as systems with constraints: \\ a geometric point of view}
	
	\author{M. F. Araujo de Resende}
	\email{resende@if.usp.br}
	\affiliation{Instituto de Física, Universidade de São Paulo, 05508-090 São Paulo SP, Brasil}
	
	\date{\today}
	
	\begin{abstract}
		In this paper, we briefly review the Hamiltonian formulation of classical systems that are constrained to submanifolds so that, within this context, the true meaning of classical gauge theories becomes clear. Please note that this paper is nothing more than a near-literal translation of Ref. \cite{mf-constraints}, which we originally published in Brazilian Portuguese in 2018. Therefore, if you, the reader, find this paper useful enough to cite it in any of your works, we kindly ask that you (also) cite Ref. \cite{mf-constraints}. \bigskip
		
	\end{abstract} 
	
	\maketitle
	
	\section{\label{intro}Introduction}
	
		According to the literature, several physical theories can be classified as \emph{gauge theories}, among which we can list some well-known examples, such as Classical Electrodynamics \cite{jackson} and the Standard Model of elementary particles \cite{ait-hey-2}. In fact, with regard to Classical Electrodynamics, for instance, it is worth mentioning that it is much more than a famous example, since it is through it that the concept of \textquotedblleft gauge theory\textquotedblright  \hspace*{0.01cm} is usually introduced to students. And this introduction is made under the assertion that the equations of motion of this theory are \emph{covariant}: i.e., under the assertion that the electric and magnetic fields do not vary when we make specific changes to the potentials that define them \cite{ait-hey-1}. Note that this occurs due to the fact that these fields are defined by derivatives of the potentials. Therefore, to keep them invariant, any transformation that is cancelled by these derivatives will be valid.
		
		Although it is not difficult to show the invariance of these fields, and consequently of the equations of motion of a gauge theory after these \textquotedblleft specific changes\textquotedblright , something that is still not very clear to some people is what a gauge theory actually is. In fact, why are gauge theories called that? Is there something more fundamental behind the fact that the equations of motion of a physical system continue to be expressed in the same format even after these changes?
		
		In view of all these possible questions, the goal of this paper is simple: to offer you, the reader, a text (which attempts to be as pedagogical as possible within the standards of a scientific paper) with a relatively simple, yet fundamental interpretation, which aims to clarify what a classical gauge theory is. We do this based on a fact that is not usually presented in the common literature on this subject: that a classical gauge theory is one that describes a physical system that is \emph{constrained} to a \emph{submanifold} that can be fixed in \emph{infinite} ways.
		
		In order to clarify to you, the reader, what this last statement means, each section of this paper will be dedicated to a specific issue. In the next section, we provide a basic introduction to the description of classical systems, by showing why it is reasonable to assume, for instance, that the spaces, where these classical systems are defined, are identified as \emph{differentiable manifolds}. Section \ref{constraints} is dedicated to presenting the theory of classical systems with constraints under the adoption of a Hamiltonian formulation, so that, in Section \ref{gauge-th}, a classical gauge theory can be defined as a special case of theories with constraints. Finally, it is worth mentioning that, in addition to \textquotedblleft sprinkling\textquotedblright \hspace*{0.01cm} these sections with examples that are useful for understanding both systems with constraints and classical gauge theories, we use Subsection \ref{example} to illustrate how this interpretation, of a classical gauge theory as a Hamiltonian system with constraints, fits a traditional example from Classical Electrodynamics. 
		
	\section{\label{manifold-section}On the description of classical systems}
	
		Whenever we need to describe a classical system that is defined in an $ n $-dimensional space, we try to do it the simplest way possible: by associating the function $ x^{j} : \left[ \ 0 \ , \ \infty \ \right) \rightarrow \mathbb{R} $ that best fits the trajectory of this system in the $ j $-th direction, where $ j = 1 , \ldots , n $. However, there are, at least, two good reasons to say that making such an adjustment is not always so simple. And the first of these can be well understood once we note that, since an $ n $-dimensional space can be parameterised by more than one coordinate system, this adjustment of functions can be done in a \emph{non-unique} way. This is exactly what happens, for instance, with a physical system that is defined on a sphere of constant radius $ R $ that is embedded in $ \mathbb{R} ^{3} $ \cite{geo-malex} \footnote{Although this Ref. \cite{geo-malex} was not originally cited in Ref. \cite{mf-constraints}, we have added it to this translation for the sake of greater clarity. \label{nota}}. After all, although it is perfectly possible to describe this system by means of a function
		\begin{equation*}
			x \left( t \right) = \left( x^{1} \left( t \right) , x^{2} \left( t \right) , x^{3} \left( t \right) \right)
		\end{equation*}
		that is based on the Cartesian perspective of $ \mathbb{R} ^{3} $, it is also perfectly possible to describe it using another function
		\begin{equation*}
			q \left( t \right) = \left( \phi \left( t \right) , \theta \left( t \right) , R \right)
		\end{equation*}
		that uses a spherical perspective.
		
		Although, in this example, it seems that the description made by using a function $ q : \left[ \ 0 \ , \ \infty \ \right) \rightarrow \mathbb{R} ^{3} $ is more convenient (since it uses fewer non-constant functions when compared to $ x : \left[ \ 0 \ , \ \infty \ \right) \rightarrow \mathbb{R} ^{3} $), it is worth noting that real physical situations are not always as simple as that of a system defined on a sphere. And, in this case, it is precisely this observation that gives rise to the second reason why we say that making such an adjustment of functions is not always easy: because \cite{mf-msc},
		\begin{itemize}
			\item on the one hand, we are not always able to see all the symmetries that a physical system possesses (which would certainly lead us to a simpler description, given that they would reduce the number of independent parameters involved in this description),
			\item on the other hand, the Lagrangian and Hamiltonian formulations that describe a system do not always lead us to a set of equations that are easier to solve, if expressed in terms of a smaller number of parameters or non-constant functions.
		\end{itemize}
		
		In addition to these two reasons, if we also want to point out a third reason, it could perfectly relate, for instance, to the differentiability of the functions that are adjusted to the trajectories of the system. However, when we assume that these trajectories can be modelled by functions that are \emph{differentiable}\footnote{This is precisely the assumption we make when, from the function that models acceleration, we obtain the equations of motion of a system.} up to an order $ k $, it becomes perfectly valid to consider that all the positions assumed by the system belong to a \emph{differentiable manifold}. In other words, it becomes perfectly valid to consider that all these positions belong to a topological space $ \mathcal{M} _{n} $ that is not necessarily flat, which has $ n $ dimensions and can be interpreted as a generalisation of a Euclidean space \cite{elon-var}.
		
		Incidentally, it is due to this whole issue of the differentiability of a manifold\footnote{For the sake of simplicity, we will occasionally omit the term \textquotedblleft differentiable\textquotedblright \hspace{0.01cm} that predicates a manifold, just as we will also omit that this differentiability only goes up to an order $ k $.} that it becomes clear that, for all \emph{curves} that can be defined through differentiable applications $ \alpha : \left( - \varepsilon , \varepsilon \right) \rightarrow \mathcal{M} _{n} $, we are also able to define vectors that are \emph{tangent} to them at each of their points \cite{man-gr}. And, as simple as it may seem, this is an observation that is very important in the context of Classical Mechanics because it shows us that, when we describe a classical system by using a Lagrangian formulation, its description is given by a function $ L : T \mathcal{M} _{n} \rightarrow \mathbb{R} $ whose domain
		\begin{equation*}
			T \mathcal{M} _{n} = \left\{ \left( x , \dot{x} \right) : x \in \mathcal{M} _{n} \ , \ \dot{x} \in T_{x} \mathcal{M} _{n} \right\}
		\end{equation*}
		is identified as a \emph{tangent bundle} \cite{man-gr}; i.e., whose domain can be identified as a set that, because it is defined as the disjoint union of \emph{all} spaces $ T_{x} \mathcal{M} _{n} $ that are tangent at each point of $ \mathcal{M} _{n} $, is able to give us information about the position $ x = \left( x^{1} , \ldots , x^{n} \right) $ and the velocity $ \dot{x} = \left( \dot{x} ^{1} , \ldots , \dot{x} ^{n} \right) $ of the system. This information is essential for the unambiguous determination of the trajectory of this system on $ \mathcal{M} _{n} $ between times $ t_{1} $ and $ t_{2} $, which can be obtained from the \emph{Euler-Lagrange} equations
		\begin{equation}
			\frac{\delta S}{\delta x^{j}} = \frac{\partial L}{\partial x^{j}} - \frac{d}{dt} \frac{\partial L}{\partial \dot{x} ^{j}} = 0 \label{euler-lagrange}
		\end{equation}
		provided that we assume that the \emph{action}
		\begin{equation}
			S = \int ^{t_{2}} _{t_{1}} L dt \label{action}
		\end{equation}
		related to the system is \emph{stationary} \cite{landau-mec}. 
		
	\section{\label{constraints}Theories of constrained systems}
	
		Anyhow, it is worth mentioning that there is a good geometric reason for highlighting the identification of the domain of $ L $ as a tangent bundle, a reason that goes beyond mathematical jargon. And in order to better understand this reason, we must take into account that the energy of a system in Lagrangian terms can be expressed as \cite{gitman}
		\begin{equation*}
			\mathcal{E} = \frac{\partial L}{\partial \dot{x} ^{\thinspace j}} \ \dot{x} ^{\thinspace j} - L \ .
		\end{equation*}
		After all, because it is from this expression that the function, which plays a leading role in the \emph{Hamiltonian} formulation of a classical system, is constructed as
		\begin{equation*}
			H = \left. \mathcal{E} \right\vert _{\dot{x} = \bar{v} \left( q , P \right) } \ ,
		\end{equation*}
		it is precisely this that shows that the domain (or \emph{phase space}) of $ H $ is a \emph{cotangent bundle} $ T^{\ast \negthickspace } \mathcal{M} _{n} $; i.e., whose domain of $ H $ is the disjoint union of \emph{all} spaces that are \emph{cotangent} at each point of $ \mathcal{M} _{n} $, since the transformation
		\begin{equation}
			P_{\thinspace j} : = \frac{\partial L}{\partial \dot{x} ^{\thinspace j}} \label{moments}
		\end{equation}
		that defines it exchanges the parameters $ \dot{x} ^{\thinspace j} $ for their \emph{duals} $ P_{\thinspace j} $.
		
		From the most basic point of view, which is related to the transition between these two formulations (and which is also found in most books on Classical Mechanics), it can be said that the only practical difference between them is that
		\begin{itemize}
			\item while in the first (which is Lagrangian), the classical system is described by a set of $ n $ equations (\ref{euler-lagrange}) containing \emph{second order} derivatives,
			\item in the second (which is Hamiltonian) we use a set of $ 2n $ equations
			\begin{equation*}
				\dot{x} ^{j} = \frac{\partial H}{\partial P_{\thinspace j}} \quad \text{and} \quad \dot{P} _{\thinspace j} = - \frac{\partial H}{\partial x^{j}}
			\end{equation*}
			equivalent to (\ref{euler-lagrange}) to describe the same dynamics, now in terms of the positions $ x^{j} $ and the \emph{moments} $ P_{j} $, but using only \emph{first-order} derivatives \cite{lemos}.
		\end{itemize}
		In other words, the equations of motion of a classical system may even become easier to solve when written in Hamiltonian formulation, but they double in size. However, from the geometric point of view that permeates these two formulations, one of the things that can also be said is that, with regard to the definition of the domains of $ L $ and $ H $, it does not matter whether we deal with one formulation or the other. After all, since we are assuming that the positions of the system belong to a manifold $ \mathcal{M} _{n} $, these two domains are always identified as two manifolds and can, therefore, be parameterised in infinite ways \cite{man-gr}. Although, for now, this information may sound like a mere \textquotedblleft mathematical curiosity\textquotedblright , it will show its full relevance a little later, more specifically when we present the definition of what a classical gauge theory is in Subsection \ref{constrained-interpretation}.
		
		Nevertheless, before we say what this \textquotedblleft relevant information\textquotedblright \hspace*{0.01cm} is, there is already an observation to be made about the transformation (\ref{moments}), which will also be very useful when such \textquotedblleft relevant information\textquotedblright \hspace*{0.01cm} is stated. And what observation is that? That there are Lagrangian functions $ L $ where, for instance, some moments may be zero. After all, since the \emph{inverse function theorem} \cite{elon-anal} tells us that, when $ P_{\thinspace a} = 0 $, it is not possible to express $ \dot {x} ^{\thinspace a} $ as a function $ v^{\thinspace a} \left( x , P \right) $, it is precisely this nullity that shows us not only that $ P_{\thinspace a} $ is a \emph{constant of motion}: this nullity also shows us that the parameters describing the system \emph{are not independent}. That is, from the same geometric point of view we used above, it is valid to assert that the physical system described by this classical theory is \emph{constrained} to a subset $ T^{\ast } \mathcal{M} _{n-m} \subset T^{\ast } \mathcal{M} _{n} $ that is, at least, defined by
		\begin{equation}
			\Phi _{1} \left( x , P \right) = P_{a} = 0 \ . \label{constraint-1}
		\end{equation}
		When this is the case, this constraint (\ref{constraint-1}) must be incorporated into the function $ H: T^{\ast } \mathcal{M} _{n} \rightarrow \mathbb{R} $ through a \emph{Lagrange multiplier} $ \lambda ^{1} $, which is now seen as a new parameter of the system, leading to a \emph{primary Hamiltonian}
		\begin{equation}
			H^{\left( 1 \right) } = H + \lambda ^{1} \Phi _{1} \label{primary}
		\end{equation}
		capable of describing the physical system only when $ \Phi _{1} = 0 $. 
		
		\subsection{\label{regularity-equations}Regularity equations}
		
			Although (\ref{constraint-1}) is a constraint that defines the physical system under consideration, as there is no guarantee that this constraint is the only one, it is essential to check whether there are other constraints defining this subset $ T^{\ast } \mathcal{M} _{n-m} $ that, in this case, should be interpreted as a \emph{submanifold}; i.e., $ T^{\ast } \mathcal{M} _{n-m} $ is a subset that can be seen as a surface\footnote{In fact, as a \emph{hypersurface} when $ n - m > 2 $.} that is \emph{immersed} in $ T^{\ast } \mathcal{M} _{n} $. Thus, by noting that the physical system already satisfies (\ref{constraint-1}), the best way to verify this is to use, as a starting point, the fact that
			\begin{equation}
				\Phi _{1} = 0 \ \Rightarrow \ \dot{\Phi } _{1} = 0 \label{consistency}
			\end{equation}
			is a valid relation on the surface in question. After all, if $ \Phi _{1} = 0 $ is a relation that is immutable by defining a part of this surface, then $ \dot{\Phi } _{1} = 0 $ will also be.
			
			\subsubsection{\label{poisson-consistence}Poisson brackets and the Dirac-Bergmann algorithm}
			
				Given this need to assess what else may arise from (\ref{consistency}), it is interesting to note that the entire dynamics of the system in question can be expressed, in a very simple way, by using what are known as \emph{Poisson brackets} \cite{landau-mec} that, for two functions $ F , G : T^{\ast } \mathcal {M} _{n} \rightarrow \mathbb{R} $, are defined by
				\begin{equation}
					\left\{ F , G \right\} = \frac{\partial F}{\partial x^{\thinspace j}} \frac{\partial G}{\partial P_{\thinspace j}} - \frac{\partial F}{\partial P_{\thinspace j}} \frac{\partial G}{\partial x^{\thinspace j}} \ . \label{func-poisson}
				\end{equation}
				This is because, in addition to allowing us to write all the equations of motion
				\begin{equation}
					\dot{z} = \left\{ z , H^{\left( 1 \right) } \right\}
				\end{equation}
				in a slightly more \textquotedblleft concise\textquotedblright \hspace*{0.01cm} way, under the consideration that $ z = \left( x , P \right) $, they also show us that any constraint $ \Phi _{1} = 0 $, which defines the submanifold $ T^{\ast } \mathcal{M} _{n-m} $ where the physical system is defined, also satisfies an equation
				\begin{equation*}
					\dot{\Phi } _{1} = \left\{ \Phi _{1} , H^{\left( 1 \right) } \right\} = 0
				\end{equation*}
				that is only valid in this submanifold. That is, if we want to assess whether the system has another constant of motion $ \Phi _{2} = 0 $ in addition to $ \Phi _{1} = 0 $, we simply calculate $ \left\{ \Phi _{1} , H^{\left( 1 \right) } \right\} $ and set the result equal to zero: if the result is an application $ \Phi _{2} : T^{\ast  } \mathcal{M} _{n} \rightarrow \mathbb{R} $ that is independent of the previous one, it must be incorporated into (\ref{primary}) so that a new round of evaluation can be performed, by using a new Hamiltonian
				\begin{equation*}
					H^{\left( 2 \right) } = H^{\left( 1 \right) } + \lambda ^{2} \Phi _{2}
				\end{equation*}
				that is slightly more complete than $ H^{\left( 1 \right) } $ with regard to the description of the system. If, at the end of this process of investigation and incorporation of constraints, the \emph{regularity equation}
				\begin{equation}
					\dot{\Phi } _{m} = \left\{ \Phi _{m} , H^{\left( m \right) } \right\} = 0 \ , \label{regularity}
				\end{equation}
				which must be valid in the submanifold $ T^{\ast } \mathcal{M} _{n-m} $, no longer leads us to any application $ \Phi _{m+1} : T^{\ast  } \mathcal{M} _{n} \rightarrow \mathbb{R} $ independent of the previous ones $ \Phi _{1} , \ldots , \Phi _{m} $, the dynamics of the system is completely defined by
				\begin{equation}
					\dot{z} = \left\{ z, H_{\mathrm{T}} \right\} \ . \label{mov-k-z}
				\end{equation}
				Here,
				\begin{equation}
					H_{\mathrm{T}} = H^{\left( m \right) } = H + \lambda ^{a} \Phi _{a} \ , \quad \text{with} \quad a = 1 , \ldots , m \ ,
				\end{equation}
				is the total Hamiltonian of the system, whose physics is restricted only to the submanifold $ T^{\ast } \mathcal{M} _{n-m} $, which is defined by
				\begin{equation}
					\Phi \left( z \right) = \left( \Phi _{1} \left( z \right) , \ldots , \Phi _{m} \left( z \right) \right) = \left( 0 , \ldots , 0 \right) \ .
				\end{equation}
				This procedure for searching for (new) constraints is known in the literature as the \emph{Dirac-Bergmann algorithm} \cite{dirac}.
		
		\subsection{\label{remark}An important remark}
		
			Although what we are about to say may seem as technical as the terms \textquotedblleft tangent bundle\textquotedblright \hspace*{0.01cm} and \textquotedblleft cotangent bundle\textquotedblright \hspace*{0.01cm} that we mentioned a few lines ago, for a good understanding of what follows it is important to note that, in accordance with P. A. M. Dirac \cite{dirac}:
			\begin{itemize}
				\item a function $ \mathcal{F} : T^{\ast } \mathcal{M} _{n} \rightarrow \mathbb{R} $ is of \emph{first class} when, for every index $ \thinspace a $, we have
				\begin{equation*}
					\left\{ \mathcal{F} , \Phi _{\thinspace a} \right\} = 0
				\end{equation*}
				in the submanifold $ T^{\ast } \mathcal{M} _{n-m} $; otherwise, $\mathcal{F} $ is a function of \emph{second class}.
			\end{itemize}
			Despite this apparent \textquotedblleft technicality\textquotedblright , the main reason we say it is that, since any $ \Phi _{\thinspace a} $ is an example of these functions, when we consider that $ \Theta $ is a square matrix of order $ m $ with elements given by
			\begin{equation*}
				\Theta _{\thinspace ab} = \left\{ \Phi _{\thinspace a} , \Phi _{b} \right\} \ ,
			\end{equation*}
			it is possible to observe an equality between the \emph{rank} $ K $ of this matrix\footnote{That is, the number $ K $ of rows or columns that are \emph{linearly independent} in $ \Theta $ \cite{hoff}.} and the number of second-class constraints that define our theory. Consequently, if $ \Theta $ is singular\footnote{That is, if $ \Theta $ is a matrix whose determinant is equal to zero.},
			\begin{itemize}
				\item this implies that $ m - K $ constraints will be first class \cite{gitman}, and
				\item the regularity equations (\ref{regularity}) also imply that
				\begin{equation}
					\dot{\Phi } _{\thinspace a} = \left\{ \Phi _{\thinspace a} , H_{\mathrm{T}} \right\} = \left\{ \Phi _{\thinspace a} , H \right\} + \left\{ \Phi _{\thinspace a} , \Phi _{\thinspace b} \right\} \lambda ^{\thinspace b} = 0 \ , \label{regular-conditions}
				\end{equation}
			\end{itemize}
			this shows us that the $ m - K $ Lagrange multipliers (which are used to implement these $ m - K $ first-class constraints on $ H_{\mathrm{T}} $) cannot be solved unequivocally. And, as will become clear shortly, it is precisely behind this non-uniqueness that the interpretation of a gauge theory rests. 
			
	\section{\label{gauge-th}Classical gauge theories}
	
		\subsection{\label{first-example}A first example}
		
			In order to introduce you, the reader, to the idea of how a classical gauge theory can be interpreted as a special case of theories of systems with constraints, we will begin by analysing a simple example: let us consider a classical system that is modelled by the Lagrangian \cite{gitman}
			\begin{equation}
				L = \frac{1}{2} \left( \dot{x} - y \right) ^{2} \ . \label{simple-example-lag}
			\end{equation}
			After all, one of the advantages of this example is that, from its equations of motion
			\begin{equation}
				\frac{\delta S}{\delta x} = \dot{y} - \ddot{x} = 0 \quad \text{and} \quad \frac{\delta S}{\delta y} = y - \dot{x} = 0 \ , \label{example-movement-equations}
			\end{equation}
			it is already clear that there is a restriction (i.e., a \emph{constraint}) between the behaviour of $ y \left( t \right) $ and $ \dot{x} \left( t \right) $.
			
			By the way, it is worth noting that this same constraining conclusion, which appears in (\ref{example-movement-equations}) under this Lagrangian frame, also appears naturally when we adopt a Hamiltonian formulation, given that, regardless of the formulation we use to analyse a system, it remains the same: after all, note that, in addition to the relations
			\begin{equation*}
				P_{x} = \frac{\partial L}{\partial \dot{x}} = \dot{x} - y \quad \text{and} \quad P_{y} = 0
			\end{equation*}
			show us that
			\begin{equation*}
				\Phi _{1} = P_{y} = 0
			\end{equation*}
			should be interpreted as the primary constraint of the system in question, it is from the primary Hamiltonian
			\begin{equation}
				H^{\left( 1 \right) } = H + \lambda ^{1} \Phi _{1} \label{primary-example}
			\end{equation}
			(which is obtained by using the prescription made in the last Section), where
			\begin{equation*}
				H = \frac{1}{2} P^{2} _{x} + y P_{x} \ ,
			\end{equation*}
			that the relation $ y - \dot{x} = 0 $ ends up \textquotedblleft reappearing\textquotedblright \hspace*{0.01cm} as a \emph{secondary} constraint of the system, given that
			\begin{equation*}
				\dot{\Phi } _{1} = \left\{ \Phi _{1} , H^{\left( 1 \right) } \right\} = \left\{ P_{y} , y \right\} P_{x} = 0 \ \Rightarrow \ \Phi _{2} = P_{x} = 0 \ .
			\end{equation*}
			
			In any case, and due to the simplicity of these results, something that you, the reader, may be wondering about is why this specific example is being used to introduce the idea of a classical gauge theory. And the best answer we can give to readers who are asking themselves this question is divided into two parts, the first of which can be easily understood as long as they note that nothing new arises from
			\begin{equation*}
				\dot{\Phi } _{2} = \left\{ \Phi _{2} , H^{\left( 2 \right) } \right\} = 0 \ .
			\end{equation*}
			That is, since this regularity equation is identically zero for a secondary Hamiltonian
			\begin{equation*}
				H^{\left( 2 \right) } = H^{\left( 1 \right) } + \lambda ^{2} \Phi _{2} = \frac{1}{2} P^{2} _{x} + \lambda _{1} P_{y} + \left( y + \lambda _{2} \right) P_{x}
			\end{equation*}
			that is slightly more \textquotedblleft refined\textquotedblright \hspace*{0.01cm} than (\ref{primary-example}), the only constraints to which the system is subject are
			\begin{equation*}
				\Phi _{1} = P_{y} = 0 \quad \text{and} \quad \Phi _{2} = P_{x} = 0 \ .
			\end{equation*}
			And the great virtue of this conclusion is not even that these are the only two constraints of the physical system: the great virtue of this conclusion is that, as
			\begin{equation*}
				\left\{ \Phi _{1} , \Phi _{2} \right\} = \left\{ P_{y} , P_{x} \right\} = 0 \ ,
			\end{equation*}
			this is exactly what characterises these two constraints as first-class.
			
			In this fashion, as we suggested, at the end of the last section, that the presence of first-class constraints is directly related to the characterisation of a theory as a \textquotedblleft gauge theory\textquotedblright , it becomes clear why we used this system, which is modelled by (\ref{simple-example-lag}), as an example, even though: it is only one among several; and we have not yet explained how this presence of first-class constraints characterises a classical gauge theory. This last point will be explained only in Subsection \ref{constrained-interpretation}.
			
			\subsubsection{\label{usual-interpretation}On the usual interpretation of a gauge theory}
			
				However, it is when we look at the equations of motion (\ref{example-movement-equations})\footnote{More specifically at the second one, since the first can be seen as a corollary of the second.} that the second part of the best answer we can give you, the reader, appears. After all, we should note that, since the dynamics of the system in question must be such that $ y - \dot{x} = 0 $, the initial boundary conditions
				\begin{equation*}
					x \left( 0 \right) = \alpha \quad , \quad \dot{x} \left( 0 \right) = y \left( 0 \right) = \beta \quad \text{and} \quad \dot{y} \left( 0 \right) = \gamma
				\end{equation*}
				(where $ \alpha $, $ \beta $ and $ \gamma $ are three real constants) lead us to the solution
				\begin{equation*}
					x \left( t \right) = \alpha + \beta t + \frac{\gamma }{2} t^{2} + \int ^{t} _{0} \varphi \left( t^{\prime } \right) dt^{\prime } \quad \text{and} \quad y \left( t \right) = \beta + \gamma t + \varphi \left( t \right)
				\end{equation*}
				as general as possible from (\ref{example-movement-equations}). Here, $ \varphi \left( t \right) $ is a function such that $ \varphi \left( t \right) = \dot{\varphi } \left( t \right) = 0 $.
				
				Given this result, and from the fact that (\ref{example-movement-equations}) also shows us that \cite{gitman}
				\begin{equation*}
					\left( \frac{\delta }{\delta x} - \frac{d}{dt} \frac{\delta }{\delta y} \right) S = 0 \ ,
				\end{equation*}
				we see that both the action $ S $ and the Lagrangian $ L $ are \emph{invariant} under the transformations
				\begin{equation}
					x \ \rightarrow \ x^{\prime } = x + f \quad \text{and} \quad y \ \rightarrow \ y^{\prime } = y + \dot{f} \label{example-gauge-transformations}
				\end{equation}
				where $ f $ is an arbitrary function of time \cite{gitman}. That is, from the traditionally known point of view, it is precisely these transformations that allow us to characterise the system modelled by (\ref{simple-example-lag}) as a classical gauge theory: because, whatever $ f \left( t \right) $ we choose to fix a solution $ x \left( t \right) $, the physics of the system continues to be described by a set of equations that remain invariant. 
				
		\subsection{\label{constrained-interpretation}Gauge theories as Hamiltonian systems with constraints}
		
			Although this system, which is modelled by (\ref{simple-example-lag}), is a good example of a classical gauge theory featuring first-class gauge fields, we have not yet explained how the presence of these gauge fields characterises such a theory. And the best way to explain to you, the reader, why this happens is to analyse the dynamics of a classical system with constraints by using (\ref{mov-k-z}) and by taking into account something that, as we already mentioned in Section \ref{constraints}, is highly relevant: that cotangent bundles are examples of manifolds \cite{elon-var}. After all, as this implies that the parameterisation $ z $ adopted for $ T^{\ast } \mathcal{M} _{n} $ is not unique, this ensures that it is possible to take another set of parameters to describe a classical system with constraints through
			\begin{equation}
				\dot{\kappa } = \left\{ \kappa \ , H^{\thinspace \prime } _{{\mathrm{T}}}  \left( \kappa \right) \right\} _{\Phi ^{\thinspace \prime } = 0} \ . \label{mov-k-eta}
			\end{equation}
			Here, $ \Phi^{\thinspace \prime } _{a} \left( \kappa \right) = 0 $ are the new constraints and $ H^{\thinspace \prime } _{{\mathrm{T}}} : T^{\ast } \mathcal{M} _{n} \rightarrow \mathbb{R} $ is the new total Hamiltonian describing the system, all adapted to this new parameterisation $ \kappa $.
			
			As a result of this line of reasoning, the recognition that $ T^{\ast } \mathcal{M} _{n-m} $ is a submanifold of $ T^{\ast } \mathcal{M} _{n} $ is directly related to the fact that the differentiable decomposition \cite{man-gr}
			\begin{equation}
				T^{\ast } _{\mathsf{q}} \mathcal{M} _{n} = T^{\ast } _{\mathsf{q}} \mathcal{M} _{n-m} \oplus \left( T^{\ast } _{\mathsf{q}} \mathcal{M} _{n-m} \right) ^{\perp } \label{decomposition}
			\end{equation}
			holds at any point $ \mathsf{q} \in \mathcal{M} _{n-m} $, where $ \left( T^{\ast } _{\mathsf{q}} \mathcal{M} _{n-m} \right) ^{\perp } $ is the orthogonal complement of $ T^{\ast } _{\mathsf{q}} \mathcal{M} _{n-m} $ \footnote{Which can also be called the \emph{normal space of immersion} at the point $ \mathsf{q} $ \cite{man-gr}.}, a set of parameters $ \kappa = \left( \omega , \Omega \right) $ ends up becoming special: the one where $ \omega = \left( q , p \right) $ and $ \Omega = \left( \mathcal{Q} , \mathcal{P} \right) $ are \emph{intrinsic} parameterisations only of $ T^{\ast } _{\mathrm{q}} \mathcal{M} _{n-m} $ and of $ \left( T^{\ast } _{\mathrm{q}} \mathcal{M} _{n-m} \right) ^{\perp } $ respectively. And the main reason why we affirm that this new parameterisation $ \kappa $ is special is that, regardless of any geometric considerations and/or interpretations, it has already been demonstrated in Ref. \cite{tyutin} that there exists a parameterisation $ \kappa ^{\prime } = \left( \omega ^{\prime } , \Omega ^{\prime } \right) $ where a new total Hamiltonian
			\begin{equation}
				H_{{\mathrm{T}}} \negthinspace \left( \kappa ^{\prime } \right) = H_{\mathrm{F}} \negthinspace \left( \omega ^{\prime } \right) + \lambda _{\mathcal{P} ^{\prime }} \mathcal{P} ^{\prime } + \mathcal{O} \bigl( \dot{\mathcal{P} ^{\prime }} , \mathcal{P} ^{\prime 2} \bigr) \label{hamiltonian-reexpress}
			\end{equation}
			for a classical system with constraints can be expressed by using
			\begin{enumerate}
				\item[(a)] a pair of canonically conjugated variables $ \omega ^{\prime } = \left( q^{\prime } , p^{\prime } \right) $ \footnote{In other words, variables such that $ \left\{ q^{\prime \mu } , p^{\prime } _{\nu } \right\} = \delta ^{\mu } _{\nu \ } $ for all $ \mu , \nu = 1 , \ldots , n-m $.} that parameterises only $ T^{\ast } \mathcal{M} _{n-m} \subset T^{\ast } \mathcal{M} _{n} $, and
				\item[(b)] another pair of variables $ \Omega ^{\prime } = \left( \mathcal{Q} ^{\prime } , \mathcal{P} ^{\prime } \right) $ that are also canonically conjugate, for which there is a bijection between the components of $ \mathcal{P} ^{\prime } = \left( \mathcal {P} ^{\prime } _{\thinspace \mathtt{I}} , \mathcal{P} ^{\prime } _{\thinspace \mathtt{II}} \right) $ and those of $ \Phi = \left( \Phi _{1} , \ldots , \Phi _{m} \right) $.
			\end{enumerate}
			In other words, by comparing these two parameterisations $ \kappa $ and $ \kappa ^{\prime } $, we can conclude that they are, in fact, the same.
			
			The interesting aspect that follows from this conclusion is that, when we develop (\ref{mov-k-eta}), the equations of motion of the classical system under consideration reduce to \cite{gitman}
			\begin{equation}
				\dot{\omega } = \left\{ \omega , H_{\mathrm{F}} \right\} \ , \quad \dot{\mathcal{Q}} _{\thinspace \text{I}} = \lambda _{\mathcal{P} _{\thinspace \text{I}}} \ , \quad \dot{\mathcal{Q}} _{\thinspace \text{II}} = \mathcal{A} \left( \omega , \mathcal{Q} \right) \quad \text{and} \quad \mathcal{P} = 0 \ , \label{eq-motion}
			\end{equation}
			where: $ H_{\mathrm{F}} : T^{\ast \negthickspace } \mathcal{M} _{n-m} \rightarrow \mathbb{R} $ is what we can call the \emph{physical Hamiltonian}; and $ \lambda _{\mathcal{P} _{\thinspace \textnormal{I}}} $ are the new Lagrange multipliers that implement the new first-class constraints $ \mathcal{P} _{\thinspace \textnormal{I}} = 0 $ to the new Hamiltonian (\ref{hamiltonian-reexpress}). And this, indeed, is the most important aspect related to the characterisation of a classical gauge theory: since the non-univocity, of the new multipliers that implement the new first-class constraints, allows us to make infinite choices to solve
			\begin{equation}
				\dot{\mathcal{Q}} _{\thinspace \text{I}} = \lambda _{\mathcal{P} _{\thinspace \text{I}}} \quad \text{and} \quad \dot{\mathcal{Q}} _{\thinspace \text{II}} = \mathcal{A} \left( \omega , \mathcal{Q} \right) \ , \label{calibre}
			\end{equation}
			is this whole aspect \textquotedblleft dismantled\textquotedblright \hspace*{0.01cm} of equations (\ref{eq-motion}) that shows that, whatever the \emph{gauge} $ \lambda _{\mathcal{P} _{\thinspace \textnormal{I}}} $ we fix to solve (\ref{calibre}), our choice will \emph{never} interfere with the solution of the physical equations
			\begin{equation}
				\dot{\omega } = \left\{ \omega , H_{\mathrm{F}} \right\} \ . \label{physical-equation}
			\end{equation}
			Therefore, since all the parameterisations we can choose for a manifold are related (to each other) through \emph{diffeomorphisms}\footnote{That is, through differentiable applications whose inverse is also differentiable.}, it is possible to assert that a classical gauge theory is much more than one that describes a physical system
			\begin{itemize}
				\item that is constrained to a submanifold $ \mathcal{M} _{n-m} \subset \mathcal{M} _{n} $,
				\item where the set of constraints (which defines this submanifold $ T^{\ast \negthickspace } \mathcal{M} _{n-m} \subset T^{\ast \negthickspace } \mathcal{M} _{n} $) is necessarily composed of first-class constraints.
			\end{itemize}
			Since the existence of a diffeomorphism between $ z $ and $ \kappa $ guarantees that fixing a multiplier $ \lambda _{P_{I}} $ in (\ref{eq-motion}) means fixing a multiplier $ \lambda ^{\textnormal{j}} $ in (\ref{mov-k-z}), this latter fixing is what ultimately defines an expression
			\begin{equation}
				\lambda ^{\bar{a}} = f_{\bar{a}} \left( z \right) \ , \label{gauge-characterisations}
			\end{equation}
			for all those Lagrange multipliers that, for instance, could not be determined in (\ref{regular-conditions}), which is perfectly mutable. That is, just as with the transformation (\ref{example-gauge-transformations}), since there are infinite choices of gauges we can take to solve (\ref{eq-motion}), there are also infinite choices we have to define (\ref{gauge-characterisations}) without ever
			\begin{itemize}
				\item affecting the solution of the physical equations of the system, and
				\item destroying the covariance of these same physical equations, since Poisson brackets are \emph{invariant} under canonical transformations \cite{mf-msc}.
			\end{itemize}
			
			By the way, it is precisely this aspect that justifies the term \textquotedblleft gauge\textquotedblright \hspace*{0.01cm} associated with these theories: because, just as, for instance, fixing the gauge of a measuring device means making the most convenient choice (among the various options available) to fix the scale that serves as a measure, fixing the gauge of a classical system with constraints also means making the best choice we can (among the various options available) to solve its equations of motion. 
			
		\subsection{\label{example}The free electromagnetic field as an example of a constrained system}
		
			As mentioned at the beginning of this paper, a good example of a gauge theory is Classical Electrodynamics: a theory that, because it is so well established as a physical theory, ended up serving as a backdrop for others to emerge, among which we can list
			\begin{itemize}
				\item \emph{Quantum Electrodynamics}, which was developed to describe the quantum behaviour of electrodynamic systems, and
				\item the \emph{Electroweak Model} and \emph{Quantum Chromodynamics}, which were aggregated to define the Standard Model of elementary particles.
			\end{itemize}
			However, there is a \textquotedblleft small\textquotedblright \hspace*{0.01cm} difference between the examples, which we have just mentioned, and the classical gauge theories presented in the last subsection: these examples are modelled by Lagrangian functions
			\begin{equation}
				L = \int _{\mathrm{V}} \mathcal{L} \ d \vec{x} \label{density-general}
			\end{equation}
			that use a \emph{density} $ \mathcal{L} $ that depends on \emph{fields} $ \phi ^{\mu } \left( x \right) = \phi ^{\mu } \left( \vec{x} , t \right) $ and their derivatives, where $ \mu = 0 , 1 , 2 , 3 $ and $ \mathrm{V} $ is a volume defined by the boundary conditions $ \delta \phi _{\mu } \left( x \right) = 0 $ \cite{mandl}.
			
			Although classical gauge theories with a Lagrangian (\ref{density-general}) can be interpreted perfectly along the same lines as in the previous subsection, the best way to understand the differences present in this new situation is to consider another simple example: one in which we have a physical system that, because it is composed of a single photon, is described by Maxwell's density
			\begin{equation}
				\mathcal{L} = - \frac{1}{4} F_{\mu \nu } F^{\mu \nu } \ . \label{electro-lag}
			\end{equation}
			Here, $ F_{\mu \nu } = \partial _{\mu } A_{\nu } - \partial _{\nu } A_{\mu } $ is the function that allows us to express the electric and magnetic fields respectively as
			\begin{equation}
				E_{j} = F_{j0} \quad \text{and} \quad H_{j} = \frac{1}{2} \varepsilon _{jkl} F^{kl} _{\perp } \ ,
			\end{equation}
			where $ \varepsilon _{jkl} $ is the \emph{Levi-Civita symbol} \cite{syn} where $ j = 1 , 2 , 3 $; i.e.,
			\begin{equation*}
				\varepsilon _{jkl} = \begin{cases}
					\hspace*{0.5cm} 1, \ \text{if} \ \left( j, k, l \right) \ \text{is equal to} \ \left( 1, 2, 3 \right) \ \text{or} \ \left( 2, 3, 1 \right) \ \text{or} \ \left( 3, 1, 2 \right) \ , \\
					\ - 1, \ \text{if} \ \left( j , k , l \right) \ \text{is equal to} \ \left( 3 , 2 , 1 \right) \ \text{or} \ \left( 2 , 1 , 3 \right) \ \text{or} \ \left( 1 , 3 , 2 \right) \ , \\
					\hspace*{0.5cm} 0, \ \text{otherwise} \ .
				\end{cases}
			\end{equation*} 
			
			\subsubsection{\label{gauge-foton-field}The presence of first-class constraints}
			
				According to (\ref{moments}), by expressing (\ref{electro-lag}) as
				\begin{equation*}
					\mathcal{L} = \frac{1}{2} \bigl( \dot{A} ^{j} - \partial _{j} A^{0} \bigr) ^{2} - \frac{1}{4} F_{jk} ^{2}
				\end{equation*}
				it can be seen that the fields corresponding to the conjugated moments of $ A^{0} $ and $ A^{j} $ are given respectively by
				\begin{equation}
					P_{0} = \frac{\delta \mathcal{L}}{\delta \dot{A} ^{0}} = 0 \quad \text{and} \quad P_{j} = \frac{\delta \mathcal{L}}{\delta \dot{A} ^{j}} = \dot{A} ^{j} - \partial _{j} A^{0} \ . \label{electromoments}
				\end{equation}
				Therefore, as this shows that it is not possible to express $ \dot {A} ^{0} $ as a function of the fields $ A = \bigl( A^{0} \left( x \right) , \vec{A} \left( x \right) \bigr) $ and/or $ P = \bigl( P_{0} \left( x \right) , \vec{P} \left( x \right) \bigr) $, it is entirely valid to interpret
				\begin{equation}
					\mathcal{I} _{1} = P_{0} = 0 \label{constraint-first}
				\end{equation}
				as a primary constraint, from which it follows that the Hamiltonian formalism of this electrodynamic system can be primarily defined by \cite{gitman}
				\begin{equation*}
					\mathcal{H} ^{\thinspace \left( 1 \right) } = \mathcal{H} + \lambda ^{1} \mathcal{I} _{1} \ ,
				\end{equation*}
				Here, $ \lambda ^{1} $ is the Lagrange multiplier that is necessary to implement this primary constraint on the density
				\begin{equation*}
					\mathcal{H} = \frac{1}{2} P^{2} _{j} - P_{j} \partial _{j} A^{0} + \frac{1}{4} F_{jk} ^{2} \ .
				\end{equation*}
				
				Certainly one of the differences that you, the reader, may have already noticed in the last few lines is that, as in (\ref{example-movement-equations}), instead of dealing with partial derivatives to define the conjugated moments, we are dealing with partial derivatives that are slightly different: these are \emph{functional partial derivatives} \cite{lemos}, since, like the action (\ref{action}), the Maxwell's density (\ref{electro-lag}) is not a function, but rather a \emph{functional}\footnote{That is, the Lagrangian density we are dealing with is, like all others (\ref{density-general}), an application whose variables are functions of other variables.}. As a result, it becomes clear that if we want to continue dealing with the entire framework that Poisson brackets offer for analysing our electrodynamic system, we need to rewrite them in terms of these functional partial derivatives. In this case, for two functionals $ \mathcal{F} $ and $ \mathcal{G} $ that depend on the conjugated fields $ \phi ^{\mu } \left( x \right) $ and $ P_{\mu } \left( x \right) $, the Poisson brackets between them are defined by
				\begin{equation}
					\left\{ \mathcal{F} , \mathcal{G} \right\} = \int \left[ \frac{\delta _{t} \mathcal{F}}{\delta \phi ^{\mu } \left( x \right) } \frac{\delta _{t} \mathcal{G}}{\delta P_{\mu } \left( x \right) } - \frac{\delta _{t} \mathcal{G}}{\delta \phi ^{\mu } \left( x \right) } \frac{\delta _{t} \mathcal{F}}{\delta P_{\mu } \left( x \right) } \right] d \vec{x} \ , \label{poisson-fields}
				\end{equation}
				where $ t $ is being used here as an index only to highlight the fact that $ \mathcal{F} $ and $ \mathcal{G} $ need to be fixed at the \emph{same} instant in time. Note that, from the point of view of parameterisations, it is possible to say that, in this new situation of a classical system described in terms of conjugated fields, these are responsible for parameterising a kind of \textquotedblleft functional manifold\textquotedblright \hspace*{0.01cm} where such a system is defined.
				
				Thus, by continuing the search for new constraints with the help of the \textquotedblleft new\textquotedblright \hspace*{0.01cm} Poisson brackets, as the primary constraint (\ref{constraint-first}) must satisfy
				\begin{equation*}
					\dot{\mathcal{I}} _{1} = \left\{ \mathcal{I} _{1} , \mathcal{H} ^{\thinspace \left( 1 \right) } \right\} = 0
				\end{equation*}
				on the \textquotedblleft functional submanifold\textquotedblright \hspace*{0.01cm} that it helps to define, we conclude that
				\begin{equation}
					\mathcal{I} _{2} = \partial _{j} P_{j} = 0 \label{constraint-sec}
				\end{equation}
				is the only additional constraint that appears. In this way, since $ \left\{ \mathcal{I} _{1} , \mathcal{I} _{2} \right\} = 0 $ we conclude that (\ref{constraint-first}) and (\ref{constraint-sec}) are two first-class constraints and that, therefore, we are indeed dealing with a classical gauge theory. 
				
			\subsubsection{\label{add}Further considerations}
			
				Although everything we have just presented is sufficient to associate such an electrodynamic system with a classical gauge theory, it is still interesting to illustrate how equations of motion, similar to (\ref{eq-motion}), can be obtained in terms of new fields $ \kappa \left( x \right) = \left( \omega \left( x \right) , \Omega \left( x \right) \right) $, where the pairs
				\begin{equation*}
					\omega \left( x \right) = \left( a \left( x \right) , \pi \left( x \right) \right) \quad \text{and} \quad \Omega \left( x \right) = \left( \mathcal{Q} \left( x \right) , \mathcal{P} \left( x \right) \right)
				\end{equation*}
				satisfy items (a) and (b) respectively. For this to be done, it is important to note not only that the equations of this free electromagnetic field are given by
				\begin{equation*}
					\dot{A} = \left\{ A , \mathcal{H} _{\thinspace \mathrm{T}} \right\} _{\mathcal{I} = 0} \quad \text{and} \quad \dot{P} = \left\{ P , \mathcal{H} _{\thinspace \mathrm{T}} \right\} _{\mathcal{I} = 0} \ ,
				\end{equation*}
				where
				\begin{equation*}
					\mathcal{H} _{\thinspace \mathrm{T}} = \mathcal{H} + \lambda ^{1} \mathcal{I} _{1} + \lambda ^{2} \mathcal{I} _{2}
				\end{equation*}
				is the total Hamiltonian that models it, but also that, by according to item (b), the sets of constraints $ \mathcal{I} = \left\{ \mathcal{I} _{1} \left( x \right) ; \mathcal{I} _{2} \left( x \right) \right\} $ and $ \mathcal{P} = \left\{ \mathcal{P} _{1} \left( x \right) ; \mathcal{P} _{2} \left( x \right) \right\} $ must be \emph{equivalent} and \emph{have the same number of elements}.
				
				Since we already know the set $ \mathcal{I} = \left\{ \mathcal{I} _{1} \left( x \right) ; \mathcal{I} _{2} \left( x \right) \right\} $, whose elements have already been obtained in (\ref{constraint-first}) and (\ref{constraint-sec}), it is not difficult to see that the simplest way to define the new set of constraints $ \mathcal{P} $ is to take its elements as
				\begin{equation}
					\mathcal{P} _{1} = P_{0} = 0 \quad \text{and} \quad \mathcal{P} _{2} = \partial _{j} P_{j} = 0 \ . \label{choice-1}
				\end{equation}
				Although this choice may not be the most interesting, it is due to it and (\ref{electromoments}) that it becomes clear that we can take $ \mathcal{Q} ^{1} = A^{0} $, and $ \mathcal {Q} ^{2} $ as a functional that depends on $ \vec{A} \left( x \right) = \left( A^{1} \left( x \right) , A^{2} \left( x \right) , A^{3} \left( x \right) \right) $: more specifically, by according to Appendix \ref{appendix}, this choice leads to a $ \mathcal {Q} ^{2} = - \Delta ^{-1} \partial _{j} A^{j} $, where $ \Delta ^{-1} $ is the inverse of the operator $ \Delta = \nabla ^{2} $.
				
				Perhaps at this point, you, the reader, may be wondering why (\ref{choice-1}) may not be the most interesting choice for defining the set $ \mathcal{P} $. And the best answer we can give to readers who are asking themselves is that, as Appendix \ref{appendix} also shows that
				\begin{equation}
					\mathcal{Q} ^{\prime } = \left( A^{0} , \partial _{j} A^{j} \right) \quad \text{and} \quad \mathcal{P} ^{\prime } = \left( P_{0} , - \Delta ^{-1} \partial _ {j} P_{j} \right) = \left( 0 , 0 \right) \label{choice-2}
				\end{equation}
				is another possible choice, it is precisely this other choice that ends up showing the great argument we can use to explain the usual characterisation of this electromagnetic system as a classical gauge theory. After all, as we know, from equations (\ref{calibre}), that the pair $ \mathcal{Q} ^{\prime } = \left( A^{0} , \partial _{j} A^{j} \right) $ can be arbitrarily fixed without ever affecting either the solution or the covariance of the equations describing the physical system, the simplest choice
				\begin{equation}
					A^{0} = 0 \quad \text{and} \quad \partial _{j} A^{j} = 0 \label{transverse}
				\end{equation}
				that we can make to solve $ \mathcal{Q} ^{\prime } $ already corresponds to the same \emph{transverse gauge} fixing \cite{transverse} that is usually adopted when no electromagnetic source is present \cite{jackson}. However, the even more interesting aspect behind this gauge fixing is related precisely to the transformation
				\begin{equation}
					A^{\mu } \ \rightarrow \ A^{\prime \mu } = A^{\mu } - \partial ^{\mu } \chi \ , \label{transformation}
				\end{equation}
				because, as $ \nabla ^{2} \chi = 0 $, this transformation (which most books present as characterising Classical Electrodynamics as a gauge theory) is precisely the transformation we can perform on the gauge fields $ A^{\mu } = \bigl( A^{0} , \vec{A} \bigr) $ in order to preserve canonical conjugation between the fields $ \mathcal{Q} ^{\prime } $ and $ \mathcal{P} ^{\prime } $ that appear in (\ref{choice-2}).
				
				In any case, and regardless of any preference for $ \Omega \left( x \right) = \left( \mathcal{Q} \left( x \right) , \mathcal{P} \left( x \right) \right) $ or $ \Omega \left( x \right) = \left( \mathcal{Q} ^{\prime } \left( x \right) , \mathcal{P} ^{\prime } \left( x \right) \right) $, it is noteworthy that the only thing missing, in order to complete the Hamiltonian formulation in terms of the fields that satisfy items (a) and (b), is to find an expression for the intrinsic fields $ \omega \left( x \right) = \left( a \left( x \right) , \pi \left( x \right) \right) $. With this proposal in mind, and being aware of the transverse aspect embedded in (\ref{transverse}), by noticing that the transverse components of $ \vec{A} $ are expressed as
				\begin{equation}
					A^{j} _{\perp } = \left( \delta ^{j} _{k} + \Delta ^{-1} \partial ^{j} \partial _{k} \right) A^{j} \ ,
				\end{equation}
				it is possible to see that $ a \left( x \right) = \left( A^{1} _{\perp } \left( x \right) , A^{2} _{\perp } \left( x \right) \right) $ can be taken as part of these intrinsic fields, since
				\begin{equation}
					\partial _{j} A^{j} _{\perp } = 0 \ \Rightarrow \ A^{3} _{\perp } = - \partial ^{-1} _{3} \left( \partial _{\mathtt{1}} A^{1} _{\perp } + \partial _{\mathtt{2}} A^{2} _{\perp } \right) \ . \label{a-perp}
				\end{equation}
				Therefore, since the field $ a \left( x \right) $ needs to be canonically conjugated to $ \pi \left( x \right) $, it becomes valid to take $ \pi \left( x \right) = \left( P^{\perp } _{1} \left( x \right) , P^{\perp } _{2} \left( x \right) \right) $ because its components
				\begin{equation}
					P^{\perp } _{j} = \left( \delta ^{k} _{j} + \Delta ^{-1} \partial ^{k} \partial _{j} \right) P_{j} \label{p-perp}
				\end{equation}
				also satisfy a relation similar to (\ref{a-perp}).
				
				Based on all these results, it is not only possible to demonstrate that equations of the physical system can be reduced to \cite{gitman}
				\begin{equation}
					\dot{A} ^{\xi } _{\perp } = \bigl\{ A^{\xi } _{\perp } , \mathcal{H} _{\mathrm{F}} \bigr\} = P^{\perp } _{\xi } \quad \text{and} \quad \dot {P} ^{\perp } _{\xi } = \bigl\{ P^{\perp } _{\xi } , \mathcal{H} _{\mathrm{F}} \bigr\} = \Delta A^{\xi } _{\perp } \ , \label{equations}
				\end{equation}
				where $ \xi = 1 , 2 $ and
				\begin{equation*}
					\mathcal{H} _{\mathrm{F}} = \frac{1}{2} \left[ \left( E^{\perp } _{\xi } \right) ^{2} + H^{2} _{\xi } \right]
				\end{equation*}
				is the physical Hamiltonian of the system, but it is also possible to assert (through a \textquotedblleft second way\textquotedblright ) that, as the solution of the other (non-physical) equations \cite{gitman}
				\begin{equation*}
					\mathcal{P} _{1} = \mathcal{P} _{2} = 0 \ , \quad \dot{\mathcal{Q}} ^{1} = \lambda _{\mathcal{P} _{\thinspace 1}} \quad \text{and} \quad \dot{\mathcal{Q}} ^{2} = - \Delta \mathcal{Q} ^ {1}
				\end{equation*}
				does not affect either the solution or the covariance of the physical equations (\ref{equations}), we are indeed dealing with a classical gauge theory. 
				
	\section{\label{conclusions}Final remarks}
	
		As should be clear from everything we have presented above, the purpose of this paper was simply to offer you, the reader, a text (which attempted to be as pedagogical as possible within the standards of a scientific paper) with a slightly more fundamental interpretation of what a classical gauge theory is, by using some geometric considerations and the Hamiltonian formulation of a classical system with constraints.
		
		There are several references on the interpretation of classical gauge theories as theories of systems with constraints. For readers interested in this subject, we can point to Refs. \cite{gitman}, \cite{teitel} and \cite{deriglazov} \footnote{By paraphrasing what we already said in footnote \ref{nota}, it is also worth noting that we only added Ref. \cite{deriglazov} in this translation for the sake of greater clarity.}, for instance, where it is possible to obtain much more specific and advanced information on everything we have presented here. By the way, and by speaking of these references, it is worth mentioning to you, the reader, that everything we have presented throughout this work can be interpreted as a fairly simple geometric analysis of what is contained, for instance, in Ref. \cite{gitman}.
		
		Once again, it is worth highlighting that this paper is nothing more than a near-literal translation of Ref. \cite{mf-constraints}, which we originally published in Brazilian Portuguese in 2018. After all, since there is a real possibility that we will cite this same Ref. \cite{mf-constraints} in future works due to its intelligibility, we thought it would be interesting to translate it into English in order to offer that same intelligibility to a wider audience, since not everyone in the world is fluent in Brazilian Portuguese. In other words, we hope that this translation will be useful in some way, especially for those seeking to understand what a gauge theory is in a more fundamental way. Therefore, if you, the reader, find this paper useful enough to, for instance, cite it in any of your works, we kindly ask that you (also) cite Ref. \cite{mf-constraints}.
		
	\section{Acknowledgements}
	
		This work was funded by CAPES (ProEx) and CNPq (process number 162117/2015-9). We would like to thank J. L. M. Assirati, A. F. Morais, S. R. A. Salinas, and the RBEF reviewer (whose name we do not know, but who evaluated this work) for all the discussions and suggestions that made this text better to read. 
		
	\appendix
	
	\section{\label{appendix}Supplementary calculations}
	
		According to the first choice (\ref{choice-1}), which gives $ \mathcal{Q} ^{1} = A^{0} $ and $ \mathcal {Q} ^{2} $ as a functional that depends exclusively on $ \vec{A} \left( x \right) = \left( A^{1} \left( x \right) , A^{2} \left( x \right) , A^{3} \left( x \right) \right) $, we can note that
		\begin{eqnarray*}
			\left\{ \mathcal{Q} ^{1} \left( \vec{x} \right) , \mathcal{Q} ^{1} \left( \vec{y} \right)  \right\} & = & \left\{ \mathcal{Q} ^{1} \left( \vec{x} \right) , \mathcal {Q} ^{2} \left( \vec{y} \right) \right\} = \left\{ \mathcal{Q} ^{2} \left( \vec{x} \right) , \mathcal{Q} ^{2} \left( \vec{y} \right)  \right\} \\
			& = & \left\{ \mathcal{Q} ^{1} \left( \vec {x} \right) , \mathcal{P} _{2} \left( \vec{y} \right)  \right\} = \left\{ \mathcal{Q} ^{1} \left( \vec{x} \right) , \mathcal{P} _{2} \left( \vec{y} \right)  \right\} = 0 \ ,
		\end{eqnarray*}
		as well as
		\begin{equation}
			\left\{ \mathcal{Q} ^{1} \left( \vec{x} \right) , \mathcal{P} _{1} \left( \vec{y} \right) \right\} = \delta \left( \vec{x} - \vec{y} \right) \ . \label{canonical-condition}
		\end{equation}
		Thus, since
		\begin{eqnarray*}
			\left\{ \mathcal{Q} ^{2} \left( \vec{x} \right) , \mathcal{P} _{2} \left( \vec{y} \right)  \right\} & = & \left\{ \mathcal{Q} ^{2} \left( \vec{x} \right) , \partial _{j} p_{j} \left( \vec{y} \right) \right\} \\
			& = &  \int \frac{\delta \mathcal{Q} ^{2} \left( \vec{x} \right) }{\delta A^{\mu } \left( \vec{x} ^{\ \prime } \right) } \frac{\delta }{\delta p_{\mu } \left( \vec{x} ^{\ \prime } \right) } \left[ \frac{\partial _{j} p_{j} \left( \vec{y} \right) }{\partial y^{j}} \right] d \vec{x} ^{\ \prime } \\
			& - & \int \frac{\delta }{\delta A^{\mu } \left( \vec{x} ^{\ \prime } \right) } \left[ \frac{\partial _{j} p_{j} \left( \vec{y} \right) }{\partial y^{j}} \right] \frac{\delta \mathcal{Q} ^{2} \left( \vec{x} \right) }{\delta p_{\mu } \left( \vec{x} ^{\ \prime } \right) } d \vec{x} ^{\ \prime } \\
			& = & - \frac{\partial }{\partial y^{j}} \left[\frac{\delta \mathcal{Q} ^{2} \left( \vec{x} \right) }{\delta A^{j} \left( \vec{y} \right) } \right] 
		\end{eqnarray*}
		must also be equal to $ \delta \left( \vec{x} - \vec{y} \right) $, when we note that
		\begin{equation*}
			\frac{\delta A^{j} \left( \vec{x} \right) }{\delta A^{j} \left( \vec{y} \right) } = \delta \left( \vec{x} - \vec{y} \right) \ ,
		\end{equation*}
		it is easy to conclude that
		\begin{eqnarray*}
			\left\{ \mathcal{Q} ^{2} \left( \vec{x} \right) , \mathcal{P} _{2} \left( \vec{y} \right)  \right\} & = & \delta \left( \vec{x} - \vec{y} \right) \\
			& & \Leftrightarrow \ \frac{\delta \mathcal{Q} ^{2} \left( \vec{x} \right) }{\delta A^{j} \left( \vec{y} \right) } = - \frac{\delta }{\delta A^{j} \left( \vec{y} \right) } \left\{ \Delta ^{-1} \left[ \frac{\partial A^{j} \left( \vec{x} \right) }{\partial x^{j}} \right] \right\}
		\end{eqnarray*}
		and, therefore, that $ \mathcal{Q} ^{2} = - \Delta ^{-1} \partial _{j} A^{j} $. In the case of the second choice (\ref{choice-2}), it can be obtained by taking the transformation $ A^{\prime \mu } = P_{\mu } $ and $ P^{\prime } _{\mu } = - A^{\mu } $, which leads us to another pair
		\begin{equation*}
			\mathcal{Q} ^{2 \prime } = \partial _{j} A^{j} \quad \text{and} \quad \mathcal{P} ^{\prime } _{2} = - \Delta ^{-1} \partial _{j} P_{j}
		\end{equation*}
		with canonically conjugated fields.

	\bigskip{\small \smallskip\noindent Updated: \today.}


\begin{thebibliography}{50}
		\bibitem{mf-constraints} M. F. Araujo de Resende: \emph{Rev. Bras. Ensino Fís.} \textbf{40} (1), e1312 (2018). 
		\bibitem{jackson} J. D. Jackson: \emph{Classical Electrodynamics}, Third Edition (John Willey \& Sons Inc., New York  1999).
		\bibitem{ait-hey-2} I. J. R. Aitchison, A. J. G Hey: \emph{Gauge Theories in Particles Physics, Volume II: Non-Abelian Gauge Theories: QCD and the Electroweak Theory} (Institute of Physics Publishing, Bristol and Phyladelphia 2004).
		\bibitem{ait-hey-1} I. J. R. Aitchison, A. J. G. Hey: \emph{Gauge Theories in Particles Physics, Volume I: From Relativistic Quantum Mechanics to QCD} (Institute of Physics Publishing, Bristol and Phyladelphia 2003).
		\bibitem{geo-malex} M. M. Alexandrino: \emph{Introdução à Geometria Riemanniana} (IME-USP Lecture Notes, São Paulo 2022).
		\bibitem{mf-msc} M. F. A. de Resende: \emph{Quantização da partícula não relativística em espaços curvos como superfícies do $ \mathbb{R} ^{n} $} (Dissertação de Mestrado IFUSP, São Paulo 2011).
		\bibitem{elon-var} E. L. Lima: \emph{Variedades Diferenciáveis} (Monografia de Matemática, IMPA, Rio de Janeiro 1973).
		\bibitem{man-gr} M. P. do Carmo: \emph{Riemannian Geometry} (Birkh\"{a}user, Boston 1993).
		\bibitem{landau-mec} L. D. Landau, E. M. Lifshitz: \emph{Mechanics} (Pergamon Press, New York 1976).
		\bibitem{gitman} D. M. Gitman, I. V. Tyutin: \emph{Quantization of Fields with Constraints} (Springer-Verlag, Berlin Heidelberg 1990).
		\bibitem{lemos} N. A. Lemos: \emph{Mecânica Analítica -- Segunda Edição} (Editora Livraria da Física, São Paulo 2007).
		\bibitem{elon-anal} E. L. Lima: \emph{Análise Real, Volume II} (Coleção Matemática Universitária IMPA, Rio de Janeiro 2007).
		\bibitem{dirac} P. A. M. Dirac: \emph{Lectures on Quantum Mechanics} (Yeshiva University Press, New York 1964).
		\bibitem{hoff} K. Hoffman, R. A. Kunze: \emph{Linear Algebra} (Prentice-Hall, New Delhi 1967).
		\bibitem{tyutin} D. M. Gitman, I. V. Tyutin, Y. S. Prager: \emph{Soviet Phys. Journ.} \textbf{26} 760 (1983).
		\bibitem{mandl} F. Mandl, G. Shaw G: \emph{Quantum Field Theory -- 2nd Edition} (John Wiley \& Sons Ltd., West Sussex 2010).
		\bibitem{syn} J. L. Synge, A. Shild: \emph{Tensor Calculus} (Dover Publications Inc., New York 1978).
		\bibitem{transverse} L. V. Lorenz: \emph{Phil. Mag. Scr.} 3 \textbf{34} 287 (1867).
		\bibitem{teitel} M. Henneaux, C. Teitelboim: \emph{Quantization of Gauge Systems} (Princeton University Press, New Jersey 1992).
		\bibitem{deriglazov} A. Deriglazov: \emph{Classical Mechanics: Hamiltonian and Lagrangian Formalism} (Springer-Verlag, Berlin Heidelberg 2010).
	\end{thebibliography}
\end{document}